\documentclass[3p,times,procedia]{elsarticle}
\usepackage{ecrc}
\usepackage{array}
\usepackage{multirow}
\usepackage{setspace}
\usepackage{xspace}

\volume{00}

\firstpage{1}

\journalname{Nuclear Physics A}

\runauth{}

\jid{nupha}

\jnltitlelogo{Nuclear Physics A}

\usepackage{amssymb}

\usepackage[figuresright]{rotating}

\newcommand{\be}{\begin{equation}}
\newcommand{\ee}{\end{equation}}

\newcommand{\that}{{\hat t}}
\newcommand{\shat}{{\hat s}}
\newcommand{\uhat}{{\hat u}}
\newcommand{\tlt}{\bar{t}}
\newcommand{\tls}{\bar{s}}
\newcommand{\tlu}{\bar{u}}
\newcommand{\qbar}{{\bar q}}

\begin{document}

\begin{frontmatter}



\dochead{}

\title{Factorization for forward dijets in dilute-dense collisions}


\author{Elena Petreska$^{1,2}$
}
\address{$^1${\small\it Centre de Physique Th\'eorique, \'Ecole Polytechnique, }{\small\it CNRS, 91128 Palaiseau, France}\\
$^2${\small\it Departamento de F\'isica de Part\'iculas and IGFAE, Universidade de Santiago de Compostela, 15782 Santiago de Compostela, Spain} }

\begin{abstract}
We propose a factorization formula for the cross section for forward dijet production in dilute-dense collisions. The new formula is applicable for an arbitrary value of the momentum imbalance of the two jets, $k_t$. It unifies the previously derived transverse momentum dependent (TMD) factorization for small $k_t$  (of the order of the saturation scale), and the High Energy Factorization (HEF) for large $k_t$ (of the order of the momentum of the jets).
We extend the previous TMD formula, first to finite $N_c$, and then to all ranges of $k_t$ by including off-shell matrix elements. 
We present previously unpublished analytical expressions for the TMD gluon distributions in the Golec-Biernat-Wusthoff model, and their perturbative behaviour in the McLerran-Venugopalan model. 
In addition, we show directly the equivalence of the HEF and the color glass condensate formulae in the dilute target approximation. 

\end{abstract}


\end{frontmatter}


\section{Introduction}
\label{Introduction}
In ultra-relativistic proton-proton and proton-nucleus collisions, dijet production in the projectile fragmentation region is sensitive to the small-$x$ content of the target, and the large-$x$ partons in the projectile. The wave function of the projectile is obtained from perturbative quantum chromodynamics (pQCD), while the gluon distribution in the target is in the non-linear saturation regime, which is less understood. Dijet production in asymmetric collisions of a dilute projectile and dense target have been studied in different theoretical frameworks, depending on the ordering of the momentum scales present in the problem.
There are three momentum scales involved in this process, the transverse momentum of the produced jets, $P_t$, the momentum imbalance of the jets (or equivalently the transverse momentum of the incoming gluon from the target), $k_t$, and the saturation momentum, $Q_s$.  While the color glass condensate (CGC) approach~\cite{Gelis:2010nm} does not assume any particular ordering of these scales, in the most general case there is no $k_t$ factorization formula for the cross section. Factorization formulae have been previously written for particular regions of $k_t$ values. Namely, the transverse momentum dependent (TMD) factorization~\cite{Bomhof:2006dp} is valid when the momentum imbalance is close to $Q_s$, and both are much smaller than the hard momentum of the jets, $k_t\sim Q_s\ll P_t$. On the other hand, the high-energy factorization (HEF)~\cite{Catani:1990eg}~\cite{Deak:2009xt} describes the region of large $k_t$ values on the order of $P_t$, $Q_s\ll k_t\sim P_t$. We make a connection between the different formalisms, and derive a unifying formula valid for the whole range of $k_t$ values between $Q_s$ and $P_t$~\cite{Kotko:2015ura}.



\section{Direct connection between CGC and HEF in the dilute target limit}
\label{Result1}

In the HEF formula for dijet production the dilute projectile is represented with a parton distribution function of collinear factorization, $f_{a/p}(x_1,\mu^2)$, the dense target with one $k_t$-dependent gluon distribution, ${\cal F}_{g/A}(x_2,k_t)$, and the hard part of the scattering with off-shell matrix elements, $\overline{{\cal M}_{ag^*\to cd}}$~\cite{Deak:2009xt},~\cite{Kutak:2012rf}.
%
%
The HEF formula is an ansatz which turns out to be valid for large $k_t$ values, with ${\cal F}_{g/A}(x_2,k_t)$ equal to the unintegrated gluon distribution entering in the cross section for deep inelastic scattering (DIS). This distribution is simply related to the quark-anitquark dipole scattering $S$-matrix, 
%
%
$S^{(2)}_{q\bar{q}}$, which in turn is a correlator of two fundamental Wilson lines, $S^{(2)}_{q\bar{q}}({\bf{v}},{\bf{v'}})=1/N_c \left<{\textnormal {Tr}}\, {U}({\bf{v}}){U}^\dagger({\bf{v'}})\right>$.

In the CGC, from a first-principles calculation, 
the cross section 
involves four-point and three-point correlators of fundamental and/or adjoint Wilson lines, describing the multiple scatterings of a quark and a gluon in the amplitude and its complex conjugate. 
%
%
%
%
Generally, these correlators cannot be reduced to the dipole $S^{(2)}_{q\bar{q}}$. However, if one considers a target with two scattering centres (the dilute target approximation), this reduction is possible, and all of the correlators can be related to the gluon distribution ${\cal F}_{g/A}(x_2,k_t)$ from the HEF formula. This approximation amounts to expanding the Wilson lines to second order in the background field of the target.
In momentum space, this is equivalent to taking the limit $k_t \gg Q_s$, which coincides with the region of validity of the HEF formula. The cross section 
for the ${qg^*\to qg}$ channel, for example, was calculated in Ref.~\cite{Marquet:2007vb}, and it was shown that it involves a four-point correlator of two fundamental Wilson lines, $U$, and two adjoint Wilson lines, $V$, (all at different transverse coordinates) $S^{(4)}_{qg\bar{q}g} \sim \left<{\textnormal {Tr}} (U U^\dagger t^dt^c) (V V^\dagger)^{cd} \right>$, and a three-point correlator of two fundamental and one adjoint Wilson line, $S^{(3)}_{qg\bar{q}} \sim \left<{\textnormal {Tr}} (U^\dagger t^c U t^d)V^{cd} \right>$. The four- and three-point correlators, in the approximation described above, are:
\begin{eqnarray}
\hspace{-0.7cm} &&S^{(4)}_{qg\bar{q}g}({\bf{b}},{\bf{x}},{\bf{b'}},{\bf{x'}}) \simeq   S^{(2)}({\bf{b}},{\bf{b'}})
-\frac{C_A}{C_F}\left[1- S^{(2)}({\bf{x}},{\bf{x'}})\right] - \frac{C_A}{2C_F}\left[S^{(2)}({\bf{x'}},{\bf{b}})+S^{(2)}({\bf{x}},{\bf{b'}}) - S^{(2)}({\bf{x}},{\bf{b}})-S^{(2)}({\bf{x'}},{\bf{b'}})\right], \nonumber \\
\hspace{-0.7cm} &&S^{(3)}_{qg\bar{q}} ({\bf{b}},{\bf{x}},{\bf{v'}}) \simeq  \frac{C_A}{2C_F}\left[S^{(2)}({\bf{b}},{\bf{x}})+S^{(2)}({\bf{x}},{\bf{v'}})-\frac{1}{C_A^2}S^{(2)}({\bf{b}},{\bf{v'}})\right]- \frac{C_A}{2C_F}\, .
\end{eqnarray}
In this way, we rewrite all multi-point correlators of Wilson lines in the CGC dijet cross section only in terms of the DIS distribution ${\cal F}_{g/A}(x_2,k_t)$, while reproducing the matrix elements $\overline{{\cal M}_{ag^*\to cd}}$ from the HEF formula exactly~\cite{Kotko:2015ura}. The matrix elements have been calculated in Refs.~\cite{Deak:2009xt},~\cite{vanHameren:2012uj} and~\cite{vanHameren:2013}. We show this equivalence of CGC and HEF for a dilute target for all three channels, ${qg^*\to qg}$, ${gg^*\to q\bar{q}}$ and ${gg^*\to gg}$~\cite{Kotko:2015ura}. 
\section{Unified factorization for finite $N_c$}
\label{Result2}
For the second regime of $k_t$ values, $k_t\ll P_t$ (nearly back-to-back jets), the TMD factorization formula for forward dijet production was derived in Ref.~\cite{Dominguez:2011wm}, in the large-$N_c$ limit. In contrast with the HEF formula, the part of the factorization that describes the target involves five $k_t$ dependent gluon distributions, instead of one. The $k_t$ dependence, however, is not present in the matrix elements. On the level of the hard sub-process, the incoming gluon from the target is put on shell. 

The different TMD distributions in the cross section represent the resummation of collinear gluons from the target that couple to the hard part. The resummation depends on the color flow in the $2\to2$ sub-process, and brings different gauge link structure in the gluon densities for different $2\to2$ Feynman diagrams. The gauge links, products of such resummation, turn a generic correlator of gluon field strength tensors, $\langle A|\textnormal{Tr}[
F^{i-}(\xi^+,{\bf{\xi}})F^{i-}( 0)
]|A\rangle $, into several (different) gauge invariant TMD gluon distributions that will emerge in the factorized cross section. The TMD's for all types of $2\to2$ diagrams have been calculated in Ref.~\cite{Bomhof:2006dp}.

We first generalize the derivation of the TMD formula to finite $N_c$, and then extend its validity to any value of $k_t$ between $Q_s$ and $P_t$. 
The finite-$N_c$ corrections bring three new TMD distributions, as well as corrections to the hard parts that were previously omitted. The TMD gluon distributions read \footnote{With $\int$ we denote the Fourier transform $\int \frac{2d\xi^+d^2{\bf{\xi}}}{(2\pi )^{3}p_A^{-}}e^{ix_2p_A^{-}\xi ^{+}-ik_t\cdot{\bf{\xi}}}$, where $p_A$ is the momentum of the nucleus. The gauge links are defined as $\mathcal{U}^{\left[ \pm\right] }= U(0,\pm\infty;{\bf{0}})U(\pm\infty,\xi^+;{\bf{\xi}})$, and $\mathcal{U}^{\left[\square \right] }=\mathcal{U}^{\left[ +\right] }\mathcal{U}^{\left[ -\right]\dagger}=\mathcal{U}^{\left[ -\right] }\mathcal{U}^{\left[ +\right]\dagger}$, where  $\mathcal{U}(a,b;{\bf{x}})=\mathcal{P} \exp [ ig \int_a^b dx^+ A_a^-(x^+, {\bf{x}}) t^a ]$.}:

\begin{eqnarray}
&& \hspace{-0.7cm} \mathcal{F}_{qg}^{(1)} =\int 
\left\langle \textnormal{Tr}\left[ F\left( \xi \right) \mathcal{U}^{\left[ -\right] \dagger }F\left( 0\right) \mathcal{U}^{\left[ +\right] }\right] \right\rangle\ , ~~
\label{eq:Fqg1-def}
\mathcal{F}_{qg}^{(2)} =\int 
\left\langle\textnormal{Tr}\left[ F\left( \xi \right) \frac{\textnormal{Tr}\left[ \mathcal{U}^{\left[\square \right] }\right] }{N_{c}}\mathcal{U}^{\left[ +\right] \dagger}
 F\left( 0\right) \mathcal{U}^{\left[ +\right] }\right] \right\rangle\ .\nonumber \\
&& \hspace{-0.7cm} \mathcal{F}_{gg}^{(1)} =\int 
\left\langle \textnormal{Tr}\left[ F\left( \xi \right)\frac{\textnormal{Tr}\left[ \mathcal{U}^{\left[\square \right] }\right] }{N_{c}} \mathcal{U}^{\left[ -\right] \dagger }
F\left( 0\right) \mathcal{U}^{\left[ +\right] }\right] \right\rangle\ , ~~
\label{eq:Fgg2}
\mathcal{F}_{gg}^{(2)} =\int
\frac{1}{N_c}\left\langle\textrm{Tr}\left[ F\left( \xi \right) \mathcal{U}^{\left[\square\right]\dagger} \right]
\textrm{Tr}\left[ F\left( 0\right) \mathcal{U}^{\left[ \square\right] }\right] \right\rangle ,\nonumber \\
\label{eq:Fgg3}
&& \hspace{-0.7cm} \mathcal{F}_{gg}^{(3)} =\int 
\left\langle \textnormal{Tr}\left[F\left( \xi \right) \mathcal{U}^{\left[+\right] \dagger }F\left( 0\right) \mathcal{U}^{\left[ +\right] }\right] \right\rangle
, ~~
\mathcal{F}_{gg}^{(4)} =\int 
\left\langle \textnormal{Tr}\left[F\left( \xi \right) \mathcal{U}^{\left[-\right] \dagger }F\left( 0\right) \mathcal{U}^{\left[ -\right] }\right] \right\rangle\,, \\
\label{eq:Fgg4}
&& \hspace{-0.7cm} \mathcal{F}_{gg}^{(5)} =\int 
\left\langle \textnormal{Tr}\left[F\left( \xi \right) \mathcal{U}^{\left[\square \right] \dagger } \mathcal{U}^{\left[+\right] \dagger }
F\left( 0\right) \mathcal{U}^{\left[\square \right] } \mathcal{U}^{\left[ +\right] }\right] \right\rangle\,, 
\label{eq:Fgg5}
\mathcal{F}_{gg}^{(6)} =\int 
\left\langle \textnormal{Tr}\left[ F\left( \xi \right)\mathcal{U}^{\left[ +\right] \dagger }F\left( 0\right) \mathcal{U}^{\left[ +\right] }\right]
\left(\frac{\textnormal{Tr}\left[ \mathcal{U}^{\left[\square \right] }\right] }{N_{c}} \right)^2 
\right\rangle. \nonumber
\label{eq:Fgg6}
\end{eqnarray}
The new distributions are $\mathcal{F}_{gg}^{(3-5)}$. The gluon distribution $\mathcal{F}_{qg}^{(1)}$ is the \emph{dipole distribution}, and $\mathcal{F}_{gg}^{(3)}$ is the \emph{Weizs{\"a}cker-Williams gluon distribution}.

The matrix elements accompanying these distributions are not all independent. We reduce the number of independent  matrix elements and their corresponding distributions to two per channel, and we write a more compact TMD factorization formula for forward dijet production at finite $N_c$~\cite{Kotko:2015ura} \footnote{In Eq.~(\ref{eq:gg2gg-mod}) $s$ is the center of mass energy squared, $x_1$ and $x_2$ are the longitudinal momentum fractions of the parton from the projectile and the gluon from the target, respectively, and $y_1$ and $y_2$ are their rapidities.}:
\begin{equation}
\frac{d\sigma^{pA\rightarrow {\rm dijets}+X}}{d^{2}P_{t}d^{2}k_{t}dy_{1}dy_{2}}=\frac{\alpha_{s}^{2}}{(x_1 x_2 s)^{2}}
\sum_{a,c,d} x_{1}f_{a/p}(x_{1}, \mu^2)\sum_{i=1}^{2}K_{ag\to cd}^{(i)}\Phi_{ag\rightarrow cd}^{(i)}(k_t)\ \frac{1}{1+\delta_{cd}}\ .
\label{eq:gg2gg-mod}
\end{equation}
The new hard factors, $K_{ag\to cd}^{(i)}$, are given in Table~\ref{tab:Khardfactors} in the first two columns. The respective TMD's are linear combinations of $\mathcal{F}_{ag}^{(i)}$. For the $qg \to qg$ channel $\Phi^{\left(1\right)}=\mathcal{F}_{qg}^{\left(1\right)}$ and $\Phi^{\left(2\right)}=(-\mathcal{F}_{qg}^{\left(1\right)}+N_{c}^{2}\mathcal{F}_{qg}^{\left(2\right)})/(N_{c}^{2}-1)$. 
For the $gg \to q\qbar$ channel $\Phi^{\left(1\right)}=\left(N_{c}^{2}\mathcal{F}_{gg}^{\left(1\right)}-\mathcal{F}_{gg}^{\left(3\right)}\right)
 /(N_{c}^{2}-1)$ and $\Phi^{\left(2\right)}=-N_{c}^{2}\mathcal{F}_{gg}^{\left(2\right)}+\mathcal{F}_{gg}^{\left(3\right)}$. For the $gg \to gg$ channel $\Phi^{\left(1\right)}=(N_{c}^{2}\mathcal{F}_{gg}^{\left(1\right)}-2\mathcal{F}_{gg}^{\left(3\right)}
 +\mathcal{F}_{gg}^{\left(4\right)}+\mathcal{F}_{gg}^{\left(5\right)}+N_{c}^{2}\mathcal{F}_{gg}^{\left(6\right)})
 /2N_{c}^{2}$ and $\Phi^{\left(2\right)}=(N_{c}^{2}\mathcal{F}_{gg}^{\left(2\right)}-2\mathcal{F}_{gg}^{\left(3\right)}
 +\mathcal{F}_{gg}^{\left(4\right)}+\mathcal{F}_{gg}^{\left(5\right)}
+N_{c}^{2}\mathcal{F}_{gg}^{\left(6\right)})/2N_{c}^{2}$.

\renewcommand{\arraystretch}{2.}
\begin{table}
\begin{centering}
\begin{tabular}{c|c|c||c|c}
\hline  
 & $K^{(1)}_{ag \to cd}$ &  $K^{(2)}_{ag \to cd}$& $K^{(1)}_{ag^* \to cd}(k_t)$    &$K^{(2)}_{ag^* \to cd}(k_t)$\\
\hline 
\parbox[t]{2mm}{{\rotatebox[origin=c]{90}{\small{$qg\to qg$}}}} &
$\displaystyle -\frac{C_F}{N_c} \frac{\shat (\shat^2+\uhat^2)}{ \that^2\uhat}$
&
$\displaystyle - \frac{\shat^2+\uhat^2}{2 \that^2 \shat \uhat} \left[ \uhat^2 +
\frac{\shat^2-\that^2}{N_c^2}\right]$
& $\displaystyle -\frac{\overline{u}\left(\overline{s}^{2}+\overline{u}^{2}\right)}{2\overline{t}\hat{t}\hat{s}}\left(1+\frac{\overline{s}\hat{s}-\overline{t}\hat{t}}{N_{c}^{2}\ \overline{u}\hat{u}}\right)$ & $\displaystyle -\frac{C_F}{N_c}\,\frac{\overline{s}\left(\overline{s}^{2}+\overline{u}^{2}\right)}{\overline{t}\hat{t}\hat{u}}$
\\
\hline 
\parbox[t]{2mm}{{\rotatebox[origin=c]{90}{\small{$gg\to q\bar{q}$}}}} & 
$\displaystyle \frac{1}{2 N_c} 
\frac{( \hat{t}^2 + \hat{u}^2)^2}{\hat{s}^2\hat{t}\hat{u}}$ &
$\displaystyle -\frac{1}{2C_FN_c^2} \frac{\hat{t}^2 + \hat{u}^2}{\hat{s}^2}$
&
$\displaystyle \frac{1}{2N_{c}}\,\frac{\left(\overline{t}^{2}+\overline{u}^{2}\right)\left(\overline{u}\hat{u}+\overline{t}\hat{t}\right)}{\overline{s}\hat{s}\hat{t}\hat{u}}$ & $\displaystyle \frac{\left(\overline{t}^{2}+\overline{u}^{2}\right)w}{4N_{c}^2 C_F\overline{s}\hat{s}\hat{t}\hat{u}}$
\\
\hline 
\parbox[t]{2mm}{{\rotatebox[origin=c]{90}{\small{$gg\to gg$}}}} &
$\displaystyle \frac{2 N_c}{C_F} \frac{(\hat{s}^2-\hat{t}\hat{u})^2(\hat{t}^2 +
\hat{u}^2)}{\hat{t}^2\hat{u}^2\hat{s}^2}$&
$\displaystyle \frac{2 N_c}{C_F}
\frac{(\hat{s}^2-\hat{t}\hat{u})^2}{\hat{t}\hat{u}\hat{s}^2}$
& $ \displaystyle
\frac{N_{c}}{C_F}\,\frac{v\left(\overline{u}\hat{u}+\overline{t}\hat{t}\right)}{\tlt\that\tlu\uhat\tls\shat}$ & $\displaystyle -\frac{N_{c}}{2C_F}\,\frac{v\,w}{\tlt\that\tlu\uhat\tls\shat}$
\\
\hline 
\end{tabular}
\par\end{centering}
\caption{The hard factors accompanying the gluon TMDs $\Phi_{ag\rightarrow cd}^{\left(i\right)}$, $K^{(i)}_{ag \to cd}$ for an on-shell gluon, and $K^{(i)}_{ag^* \to cd}(k_t)$ for an off-shell gluon. The Mandelstam variables are defined as $\shat = (p_1 + p_2)^2$, $\that = (p_1 - k)^2$ and $\uhat= (p_2 - k)^2$. Their bared versions are defined as $\tls= (x_2 p_{A} +p)^2$, $\tlt = (x_2 p_{A} -p_1)^2$ and $\tlu =  (x_2 p_{A} -p_2)^2$. We also denote $v=\overline{s}^{4}+\overline{t}^{4}+\overline{u}^{4}$ and $w=\overline{u}\hat{u}+\overline{t}\hat{t}-\overline{s}\hat{s}$.
\label{tab:Khardfactors}}
\end{table}
%
%

The matrix elements in the TMD factorization of Ref.~\cite{Dominguez:2011wm} were derived for an on-shell incoming gluon from the target, which limits the applicability of the formula to small values of $k_t$. We propose a solution to this limitation by including off-shell matrix elements in the factorization formula and by restoring the $k_t$ dependence in the hard part.
We calculate the $k_t$ dependent matrix elements with two methods. First, with a Feynman diagram based calculation, in the light-cone gauge for the on-shell gluons, with the gauge vector $n$ set to be equal to the four momentum of the target, $n=p_A$, and with prescribing a longitudinal polarization vector to the off-shell gluon from the target of the form $\epsilon_\mu^0=i \sqrt 2\, x_2 p_{A\, \mu}\,/|k_t|$~\cite{Catani:1990eg}. 
The second is the method of \emph{color ordered amplitudes}~\cite{Mangano:1990by}~\cite{Dixon:2013uaa}.
These \emph{dual} amplitudes represent the coefficients of a color decomposition of a generic amplitude (involving an arbitrary number of gluons and/or quarks) into a color part and a kinematic part. The color ordered amplitudes are functions of kinematic arguments only, and are gauge invariant by construction. They give the hard factors, while the color part of the decomposition, after squaring, indicates the corresponding gluon TMD. 

We apply both of the methods described above to the $2\to 2$ sub-processes involved in the dijet production at forward rapidity. The result is a factorization formula, similar to the one derived in the previous section, Eq.~(\ref{eq:gg2gg-mod}), but now with $k_t$ dependent matrix elements $K_{ag^*\to cd}^{(i)}(k_t)$,
%
%
that are given in Table~\ref{tab:Khardfactors} in the last two columns~\cite{Kotko:2015ura}. The new formula unifies the HEF approach to forward dijet production with the TMD factorization by establishing a framework applicable for hard jets, $P_t \gg Q_s$, but arbitrary $k_t$, and it is the main result of this work.

The unified factorization can be applied for pheonomenological studies of dijet azimuthal correlations in high-energy collisions. As a first step, we  study the new framework at large $N_c$, and with analytical models for the gluon distributions. We calculate the gluon densities that survive the large $N_c$ limit in the Golec-Biernat-Wusthoff model~\cite{GBW}:
\begin{eqnarray}
&&\!\!\!\!\!\!\!\!\!\!\!
 \mathcal{F}_{qg}^{(1)}(x_2, k_t)  
   =  
  2 \gamma \frac{S_\perp}{Q_s^2(x_2)} k_t^2 \exp\left[
  -\frac{k_t^2}{Q_s^2(x_2)}\right] \,, \,\, \label{FinalFqg1}
  {\cal F}_{qg}^{(2)}(x_2,k_t)
 = 
  2\gamma
  \exp\left[-\frac{k_t^{2}}{Q_s^2(x_2)}\right]
\int_1^{\infty} \frac{dt}{t(t+2)}
\exp\left[\frac{2k_t^{2}}{(t+2)Q_s^2(x_2)}\right]\,, \nonumber
  \\
  && \!\!\!\!\!\!\!\!\!\!\! {\cal F}_{gg}^{(1)}(x_2,k_t)
  = 
  \frac{\gamma}{4}
  \exp\left[-\frac{k_t^{2}}{2Q_s^2(x_2)}\right]\left(2 + \frac{k_t^{2}}{Q_s^2(x_2)} \right), \,\,\,
 {\cal F}_{gg}^{(2)}(x_2,k_t)
  = 
  \frac{\gamma}{4}
  \exp\left[-\frac{k_t^{2}}{2Q_s^2(x_2)}\right]\left(2 - \frac{k_t^{2}}{Q_s^2(x_2)} \right), \nonumber \\
&&\!\!\!\!\!\!\!\!\!\!\!  {\cal F}_{gg}^{(6)}(x_2,k_t) 
   = 
  \gamma
  \exp\left[-\frac{k_t^{2}}{2 Q_s^2(x_2)}\right]
\int_1^{\infty} \frac{dt}{t(t+1)}
\exp\left[\frac{k_t^{2}}{2(t+1)Q_s^2(x_2)}\right]\,. \label{FinalFgg6}
\end{eqnarray}
In the above expressions $\gamma = N_c S_\perp/4\pi^3 \alpha_s$, where $S_\perp$ is the transverse area of the target. We also obtain their perturbative behaviour at large $k_t$ in the McLerran-Venugopalan model~\cite{MV}. We derive the leading order term in $Q_s^2/k_t^2$, and we find that all of them scale as $ \gamma \, Q_s^2/k_t^2$, except $\mathcal{F}_{gg}^{(2)}$ that goes to zero. The above expressions for the densities, as well as numerical results that implement small-$x$ evolution, will be used for phenomenological applications of the unifying formula in a forthcoming publication. 
%
%
\section*{Acknowledgments}
\label{Acknowledgments}
EP thanks the Japan Society for the Promotion of Science (JSPS) for the travel support through the Japan-France Integrated Action Program (SAKURA). EP also thanks Quark Matter 2015 for the Fellowship for young participants.




\bibliographystyle{elsarticle-num}



\end{document}